\documentclass[twocolumn,showpacs,preprintnumbers,amsmath,amssymb]{revtex4}

\usepackage{graphicx}
\usepackage{dcolumn}
\usepackage{bm}

\begin{document}


\title{Application of the generalized two-center cluster 
model to $^{10}$Be}

\author{M. Ito$^1$}
\author{K. Kato$^2$}%
\author{K. Ikeda$^3$}
\affiliation{%
$^1$Institute of Physics, University of Tsukuba, 305-8571 Tsukuba, Japan\\
$^2$Division of Physics, Graduate School of Science, Hokkaido University,
    060-0810 Sapporo, Japan\\
$^3$RI Beam Science Laboratory, RIKEN(The institute of 
    Physical and chemical Research),
    Wako, Saitama 351-0198, Japan
}%

\date{\today}

\begin{abstract}
A generalized two-center cluster model (GTCM), including 
various partitions of the valence nucleons around two $\alpha$-cores, 
is proposed for studies on the exotic cluster structures of Be isotopes. 
This model is applied to the $^{10}$Be=$\alpha+\alpha+n+n$ system 
and the adiabatic energy surfaces for $\alpha$-$\alpha$ distances 
are calculated. It is found that this model naturally describes the 
formation of the molecular orbitals as well as that of asymptotic 
cluster states depending on their relative distance. In the 
negative-parity state, a new type of the $\alpha$+$^6$He cluster 
structure is also predicted.
\end{abstract}

\maketitle
Recent experiments by Freer {\it et al.,} \cite{FREE} revealed 
the existence of the interesting resonant states, 
which dominantly decay to  $^6$He$_{g.s.}$+$^6$He$_{g.s.}$ 
and $\alpha$+$^8$He$_{g.s.}$ 
channels in the 10 to 25 MeV excitation energy interval of $^{12}$Be. 
Similar resonant states, decaying to He-isotopes such as 
$^6$He and $^8$He, have also been observed in other Be isotopes of 
$^{10}$Be \cite{FREE} and $^{14}$Be \cite{SAITO}. 
These experiments strongly suggest the existence of exotic 
cluster states consisting of the respective He-isotope clusters.

Theoretically, Kanada-En'yo and her collaborators \cite{SURENYO} 
have studied a wide range of light nuclei within the 
anti-symmetrized molecular dynamics (AMD), and predicted many 
kinds of cluster structures in isotope chains of the 
elements Be, B and C. 
In particular, much attention has been concentrated on 
Be isotopes in which a motion of valence neutrons couples to 
the two-$\alpha$ structure of $^8$Be. 
Based on the molecular orbital model (MOM), Itagaki {\it et al.} 
has extensively studied the low-lying states of Be isotopes to clarify the 
roles of valence neutrons on a development of clusterization. 
The molecular orbitals, such as $\pi$ and 
$\sigma$ orbitals associated with the covalent binding of atomic 
molecules, have been shown to give a good description for the low-lying 
states of Be-isotopes as emphasized in the AMD studies 
\cite{ITA,ENYO}. 

On the other hand, Descouvemont and Baye \cite{DES} 
have applied a traditional cluster model 
($^6$He+$^6$He) + ($\alpha$+$^8$He), 
assuming the substructures 
of $^6$He$_{g.s.}$ and $^8$He$_{g.s.}$, to illustrate the resonances 
observed in $^{12}$Be. However, it is plausible to assume 
$^6$He and $^8$He nuclei to be stable and frozen clusters because the 
energy for two-neutrons separation is quite small in both nuclei. 
Thus, it is properly considered that the valence neutron 
of $^6$He and $^8$He are associated with companion clusters 
when they approach each other, which is similar to the 
covalent binding of valence neutrons in the low-lying states. 

The purpose of this letter is to study the molecular structures 
in low-lying and high-lying states in a unified way by taking into 
account couplings between a valence neutron's motion and a relative motion of 
the $\alpha$ clusters. 
To achieve this purpose, we propose a new framework of a generalized 
two-center cluster model (GTCM) where we can describe 
atomic-orbital motions of valence neutrons around individual 
$\alpha$ clusters on the same footing with molecular orbitals, 
being a single-particle motion around two $\alpha$-cores. 
We apply this model to the $^{10}$Be=$\alpha$+$\alpha$+$n$+$n$ 
and discuss its applicability to studies of the molecular 
structure in both the low-lying and high-lying states. 

The basis functions of GTCM for $^{10}$Be are given as 

\vspace{-2mm}
\begin{eqnarray}
\lefteqn{\Phi^{J^\pi}_K(^{10}\mbox{Be};~S)}
\nonumber\\
\nonumber\\
&&=\hat{P}^{J}_K 
{\cal A}\left\{\psi_L(\alpha)\psi_R(\alpha)
\;\sum_m\;d^{J^\pi K}_m\varphi(m)
\;\sum_n\;d^{J^\pi K}_n\varphi(n)\right\}
\nonumber\\
\nonumber\\ 
&&=\sum_{m,n}C^{J^\pi K}_{m,n}(S)\;\Phi^{J^\pi K}_{m,n}(S)\;\;,
\label{adiwf}
\end{eqnarray}

\vspace{-2mm}
\begin{eqnarray}
\lefteqn{
\Phi^{J^\pi K}_{m,n}(S)}
\nonumber\\
& &=\hat{P}^{J}_K{\cal A}\left\{\psi_L(\alpha)\;
\psi_R(\alpha)\varphi(m) 
\varphi(n)\right\} 
\equiv\hat{P}^{J}_K\hat{\Phi}^\pi_{m,n}(^{10}\mbox{Be};~S)\;,
\nonumber\\
\label{base}
\end{eqnarray}

\noindent
where $C^{J^\pi K}_{m,n}(S)$ means the product of 
$d^{J^\pi K}_m\cdot d^{J^\pi K}_n$. 
The $\alpha$-cluster wave function 
$\psi_i(\alpha)$ ($i$=$L, R$) is given by the (0s)$^4$ configuration 
in the harmonic oscillator (HO) potential 
with the relative distance-parameter $S$. The position of an 
$\alpha$-cluster is explicitly specified as the 
left (L) or right (R) side. 
A single-particle state for valence neutrons around one of $\alpha$ 
clusters is given by an atomic orbitals, 
$\varphi(i,p_n,\tau)$ with the subscripts of a center 
$i$ (=$L$ or $R$), a direction $p_n$ ($n$=$x$, $y$, $z$) of 
0$p$-orbitals and a neutron spin $\tau$ (=$\uparrow$ or $\downarrow$). 
In Eq.~(\ref{base}), the index $m$($n$) is an abbreviation of the 
atomic orbital ($i,p_n,\tau$). The basis function 
$\hat{\Phi}^\pi_{m,n}(^{10}\mbox{Be};\; S)$ 
with the parity $\pi$ is projected to the 
eigenstate of the total spin $J$ and its intrinsic angular projection $K$ 
by the projection operator $\hat{P}^{J}_K$. Various linear 
combinations of $\Phi^{J^\pi K}_{m,n}(S)$ can be shown to 
reproduce not only molecular orbital configurations but also 
cluster-model states of $^4$He+$^6$He and $^5$He+$^5$He.

First, we illustrate that the molecular-orbital configuration of 
($\sigma^+$)$^2$ can be constructed from a linear combination of 
the basis function for instance. Since $\sigma^+$ orbital is 
expressed as $\varphi(L,p_{z},\tau)-\varphi(R,p_{z},\tau)$, 
the wave function of $^{10}$Be with $(\sigma^+)^2$ of 
valence neutrons are written as follows:

\vspace{-5mm}
\begin{eqnarray}
\lefteqn{
\hat{\Phi}^+(^{10}\mbox{Be};~S)
={\cal A}\left\{ \psi_L(\alpha)\psi_R(\alpha)\right.
}\nonumber\\
& &\left.\times(\varphi(L,p_{z},\uparrow)-\varphi(R,p_{z},\uparrow)) 
\cdot
(\varphi(L,p_{z},\downarrow)-\varphi(R,p_{z},\downarrow))\right\}
\nonumber\\
\nonumber\\
& &={\cal A}\left\{ 
\psi_L(\alpha)\varphi(L,p_{z},\uparrow)\varphi(L,p_{z},\downarrow)
\;\cdot\;
\psi_R(\alpha) \right.
\nonumber\\
& & \;\;\;\;\;\;\;
-\psi_L(\alpha)\varphi(L,p_{z},\uparrow)
\;\cdot\;
\psi_R(\alpha)\varphi(R,p_{z},\downarrow) \nonumber \\
& & \;\;\;\;\;\;\;
-\psi_L(\alpha)\varphi(L,p_{z},\downarrow)
\;\cdot\;
\psi_R(\alpha)\varphi(R,p_{z},\uparrow) 
\nonumber\\
& & \;\;\;\;\;\;\;
\left. +\psi_L(\alpha)\;\cdot\;
\psi_R(\alpha)\varphi(R,p_{z},\uparrow)\;
\varphi(R,p_{z},\downarrow)\right\}.
\label{sigma}
\end{eqnarray}

\noindent
This expression means that a $^{10}$Be wave function with 
the molecular orbitals can be 
described by a linear combination, such as  
($^6$He--$\alpha$)$+$($^5$He--$^5$He)+($\alpha$--$^6$He), 
because the clusters of $^5$He(=$\alpha$+$n$) and 
$^6$He(=$\alpha$+2$n$) are given by 
${\cal A}\{\psi_i(\alpha)\varphi(i,p_n,\tau)\}$ and 
${\cal A}\{\psi_i(\alpha)\varphi(i,p_n,\tau)\varphi(i,p_{n'},\tau')\}$, 
respectively. This linear combination can be exactly constructed 
from basis functions of $\hat{\Phi}^\pi_{m,n}(^{10}\mbox{Be};\; S)$.

Next, we show that the basis functions of Eq.~(\ref{base}) 
can also describe the cluster-model states in which neutron's 
orbitals have a definite spin around one of $\alpha$-cores. 
For instance, 0$p_{3\slash2}$ $j_z$=+$3\slash2$ is 
expressed as $\left\{
\varphi(i,p_{x},\uparrow)-i\varphi(i,p_{y},\uparrow)
\right\}$. Thus, the $3\slash2^-$  states of $^5$He with the 
0$p_{3\slash2}$ neutron is written as $\sum_n{\cal A}
\left\{\psi_i(\alpha)\cdot d_n\varphi(n)\right\}$=
${\cal A}\left\{\psi_i(\alpha)\cdot\varphi(i,p_{x},\uparrow)\right\}
-{\cal A}\left\{ \psi_i(\alpha)\cdot i\varphi(i,p_{y},\uparrow)\right\}$.
Similarly, the $^6$He($I^\pi$) clusters with a definite intrinsic 
spin-parity $I^\pi$, such as [(0$p_{3\slash2}$)$^2$]$_{0^+}$, 
can be constructed from a certain linear 
combination of $\sum_{m,n}{\cal A}\{\psi_i(\alpha)\cdot d_m\varphi(m)
\;d_n\varphi(n) \}$. 
Therefore, the wave function of Eq.~(\ref{adiwf}) can describe the 
cluster-model states of [$\alpha$ $\otimes$ $^6$He($I^\pi$)] and 
[$^5$He($I_1^{\pi_1}$) $\otimes$ $^5$He($I_2^{\pi_2}$)].

The wave function of $^{10}$Be is finally given by taking a 
superposition over the relative distance-parameter $S$ and the intrinsic 
angular projection $K$ as 

\begin{eqnarray}
\Psi^{J^\pi}(^{10}\mbox{Be}) & = & \int dS\;\sum_K\;
\Phi^{J^\pi}_K(^{10}\mbox{Be};S)
\nonumber\\
&=&\int dS\; \sum_{K\beta} 
C^{J^\pi K}_{\beta}(S)\;\Phi^{J^\pi}_K(S) 
\end{eqnarray}

\noindent
with $\beta \equiv$ ($m$, $n$). The coefficients 
$C^{J^\pi K}_{\beta}(S)$ 
are determined by solving a coupled channel GCM (Generator Coordinate 
Method) equation \cite{HORI}:

\begin{eqnarray}
\lefteqn{\int dS\sum_{\beta K}C^{J^\pi K}_\beta(S)}
\nonumber\\
&&\times
\left\langle \Phi^{J^\pi K'}_{\beta'}(S')~|~H~-~E^{J^\pi}~|
~ \Phi^{J^\pi K}_\beta(S)\right\rangle=0\;\;. 
\label{GCM}
\end{eqnarray}

To see the coupling properties, we solve Eq.~(\ref{GCM}) in a step 
by step. First, we solve Eq.~(\ref{GCM}) at a fixed S. Namely, we solve 

\begin{eqnarray}
\lefteqn{\sum_{\beta K}C^{J^\pi K}_\beta(S)}\nonumber\\
&&\times\left\langle 
\Phi^{J^\pi K'}_{\beta^\prime}(S)~|~H~-~E^{J^\pi}(S)~|~ 
\Phi^{J^\pi K}_\beta(S)\right\rangle=0\;\;. 
\label{GCM2}
\end{eqnarray}

\begin{figure}
  \includegraphics[width=5.7cm]{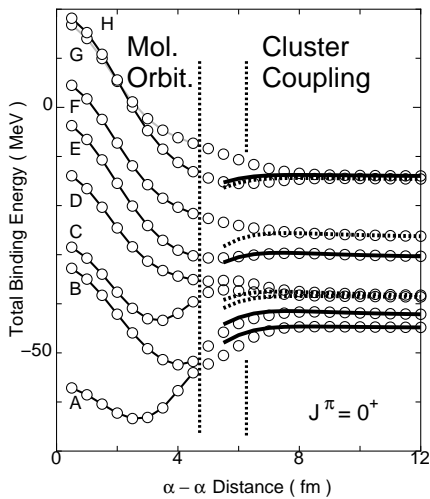} 
\vspace{-2mm}
\caption{Dynamical transition from the molecular orbitals to the 
cluster-coupling states in the adiabatic energy surfaces. 
In the right part, the solid and dotted curves show the energies of 
the $\alpha$+$^6$He and $^5$He+$^5$He cluster states, respectively.  
In the left part, the surfaces of A, B, C and D have 
a dominant component of 
($\pi_{3\slash2}^-$)$^2$, ($\sigma_{1\slash2}^+$)$^2$, 
($\pi_{1\slash2}^-$)$^2$ and ($\pi_{3\slash2}^+$)$^2$, respectively, 
while E, F, G and H ($\pi_{1\slash2}^+\sigma_{1\slash2}^+$) and 
($\pi_{1\slash2}^-\sigma_{1\slash2}^-$), 
($\sigma_{1\slash2}^-$)$^2$ and 
($\pi_{1\slash2}^+$)$^2$, respectively. 
}
\label{0+sur}
\end{figure}

\noindent
The eigenvalue $E^{J^\pi}(S)$ is a function 
of the relative distance-parameter $S$, and then we call the solutions of 
energies and wave functions 
``adiabatic energy surfaces'' and ``adiabatic eigenstates'', 
respectively. 
The calculated adiabatic energy surfaces for the 
$J^\pi$=0$^+$ state are shown by open circles in 
Fig.~\ref{0+sur}. Here, for the nucleon--nucleon interaction, 
we adopted the Volkov No.2 with the Majorana parameter $m$=0.576 and 
without the Bartlett and Heisenberg exchanges. Due to the reduction of 
the majorana parameter, the total binding energy is gained and it becomes 
easy to see the continuum states. We also employed the G3RS 
interactions for the spin-orbit parts. The radius parameter 
$b$ of HO wave functions for $\alpha$ clusters and valence neutrons is 
taken as 1.44 fm. 

At the asymptotic distance ($S$$\rightarrow$$\infty$), 
where two $\alpha$-cores are completely separated, 
we can define the asymptotic channels such as 
[$^4$He+$^6$He($I$)]$_{L}$ and [$^5$He($I_1$)+$^5$He($I_2$)]$_{IL}$, in 
which individual clusters have intrinsic spins 
($\mathbf{I_1}$, $\mathbf{I_2}$) and coupled 
with the channel spin $I$ ($\mathbf{I}$=$\mathbf{I_1 + I_2}$) 
and the relative one $L$. We call the coupling scheme of these 
asymptotic channels as ``a cluster-coupling scheme''. 
The solid and dotted curves shown in the right part of Fig.~\ref{0+sur} 
are the expectation values $<H>$ of the 
[$^4$He+$^6$He($I$)]$_{L}$ and [$^5$He($I_1$)+$^5$He($I_2$)]$_{IL}$ 
cluster-coupling schemes, respectively

From Fig.~\ref{0+sur}, we can see that in $S$ $\geq$ 6 fm, 
the calculated energy surfaces are completely the same as those of 
cluster-coupling wave functions. This means that, in this region,  
the valence neutrons are localized at one of $\alpha$-cores 
and rotation of two clusters is de-coupled to each other. 
At the asymptotic region, therefore, each nucleus keeps its 
isolated states and weakly coupled to each other. On the other hand, 
in $S\leq$ 6 fm, the energies for the 
cluster-coupling scheme deviate from the adiabatic 
energy surfaces. 

We study the adiabatic eigenstates in an internal region ($S\lesssim 4$ fm) 
in the view of the molecular orbital formation. 
As shown in an example of $(\sigma^+)^2$ in Eq.~(\ref{sigma}), 
the wave function of molecular orbitals is expressed by a 
linear combination of different kinds of cluster-wave functions, 
where each cluster has no good angular momenta.
In the molecular orbitals, the valence neutrons are moving around 
two $\alpha$-cores
with a specific direction in respect to the $\alpha$--$\alpha$ axis. 
Such a configuration of the system is called as 
``a strong-coupling scheme''. The overlap of the adiabatic eigenstates 
with the wave functions of molecular orbitals identifies the dominant 
components in the adiabatic eigenstates as shown in Fig.~\ref{0+sur}. 
The adiabatic eigenstates connected by the thin-solid curves at the
internal region have the common dominant-components of various 
kinds of molecular orbitals.

\begin{table}
\caption{\label{tab:dis}
Dominant component in the eigenstates along to the lowest 
adiabatic surface A of Fig.~\ref{0+sur}.
In up-most row, the distances of the calculated eigenstates are shown. 
The dominant basis-state is shown in the second row, while
its population is done in the lowest one. The distance parameter 
of $S$=6 fm corresponds to the ``intermediate-coupling region''. 
See text for details. 
}
\begin{ruledtabular}
\begin{tabular}{lcccc}
$S$ (fm) & 2 & 4 & 6 & 8\\
\hline
Main component & ($\pi_{3\slash2}^-$)$^2$ & ($\pi_{3\slash2}^-$)$^2$ & 
Mixed & $[\alpha$+$^6$He(0$_1^+$)$]_{L=0}$ \\
Squired ampli. & 0.93 & 0.81 & $-$ & 0.99\\
\end{tabular}
\end{ruledtabular}
\end{table}

Table \ref{tab:dis} shows the dominant molecular-orbital configurations 
in the adiabatic eigenstates along to the lowest surface A of 
Fig.~\ref{0+sur}. 
The squared amplitudes listed in Table \ref{tab:dis} have the meaning of 
the probability of finding the system in a molecular orbital, because 
an orthogonality among the different molecular orbitals is good in spite 
of the anti-symmetrization effect. This is due to the fact that the 
molecular orbitals are similar to the deformed shell-model orbitals 
in the internal region. 

We can see that the eigenstate has a
dominant component of ($\pi_{3\slash2}^-$)$^2$
at $S$=2 and 4 fm.
On the other hand, at the asymptotic region ($S\sim 8$ fm), the 
molecular orbitals are strongly mixed with each other, and  
the $[\alpha+^6$He(0$_1^+$)$]_{L=0}$ 
channel dominates. 
Thus, the cluster-coupling scheme becomes a good basis state 
for describing the system at the external region, while the 
strong-coupling one at the internal region.
At an intermediate region ($S$=5$\sim$6 fm), we find that 
the dominant component of the lowest surface A is changed 
from ($\pi_{3\slash2}$)$^2$ to ($\sigma_{1\slash2}^+$)$^2$ (72 $\%$). 
At this distance, an amplitude of the 
[$\alpha$+$^6$He(0$_1^+$)]$_{L=0}$ channel in the 
cluster-coupling schemes amounts to 61$\%$. 
This means that, in $S$=5$\sim$6 fm, the eigenstate
has ``an intermediate coupling scheme'' between the strong-coupling scheme 
and the cluster-coupling one. 

\begin{figure}
  \includegraphics[width=5.7cm]{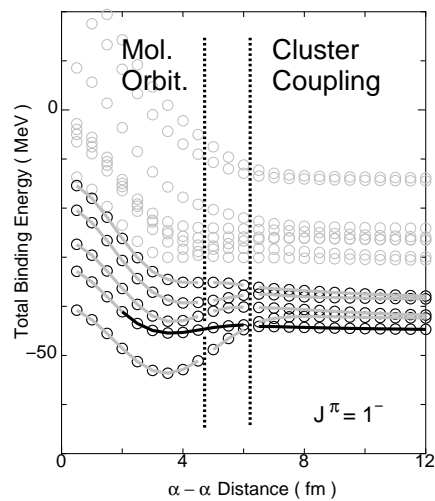}
\vspace{-2mm}
\caption{
The same as Fig.~\ref{0+sur} but for the negative parity 
states ($J^\pi$=1$^-$). In the internal region of $S\leq$4.5 fm, 
the lowest, second, third and fourth curves connected by a thin-solid 
line show the surfaces having 
a dominant component of ($\pi_{3\slash2}^-\sigma_{1\slash2}^+$)$_{K=1}$, 
($\pi_{1\slash2}^-\sigma_{1\slash2}^+$)$_{K=1}$, 
($\pi_{1\slash2}^-\sigma_{1\slash2}^+$)$_{K=0}$, 
($\pi_{3\slash2}^+\pi_{1\slash2}^-$)$_{K=0}$, respectively.
The surface with a thick curve has a dominant component of the 
[$\alpha$+$^6$He(0$_1^+$)]$_{L=1}$ channel. 
}
\label{1-sur}
\end{figure}

Finally, the GCM equation (\ref{GCM}) is solved by employing the 
basis states ranging from $S$=1 fm to $S$=9 fm with 
the mesh of 0.5 fm. We obtain the result that the lowest three 
GCM solutions have energy gains of about 
1$\sim$2 MeV, and the dominant amplitudes 
around the respective local minimums in the adiabatic energy 
surfaces A, B and C. Therefore, these solutions of the 0$_1^+$, 0$_2^+$ and 0$_3^+$ states 
are concluded to have the molecular-orbitals configuration of 
($\pi_{3\slash2}^-$)$^2$, ($\sigma_{1\slash2}^+$)$^2$ and 
($\pi_{1\slash2}^-$)$^2$, respectively \cite{ITA}.  

Similarly, we calculate the $J^\pi$=1$^-$ state. 
The calculated energy surfaces 
are shown in Fig.~\ref{1-sur}. We focus on the lowest five surfaces 
and investigated their intrinsic structure, because there is no 
definite local-minimums in higher surfaces. 
As shown by two vertical dotted-lines in Fig.~\ref{1-sur}, 
we find the dynamical transition from the strong-coupling scheme to 
the cluster-coupling one in the adiabatic surfaces. 
However, the first excited surface connected by a thick curve
has an almost pure-component of the 
[$\alpha$+$^6$He(0$_1^+$)]$_{L=1}$ channel. 
Thus, this excited surface is made by the pure cluster-coupling state 
of this channel in a wide range of the $\alpha$-$\alpha$ distance. 

The energy gain of the lowest three solution due to GCM is about 
1$\sim$3 MeV depending on the states. In the lowest and the third 
1$^-$ states, we find that the wave functions distribute around 
the respective local minimums of the adiabatic surfaces. 
Thus, the intrinsic structures of the 1$_1^-$ and 1$_3^-$ states 
are explained with the molecular orbitals of 
($\pi_{3\slash2}^-\sigma_{1\slash2}^+$)$_{K=1}$ and 
($\pi_{1\slash2}^-\sigma_{1\slash2}^+$)$_{K=1}$, 
respectively. The 1$_2^-$ state has the main components 
of the first excited surface in 
Fig.~\ref{1-sur}, which is 
interpreted in terms of the cluster-coupling states of 
[$\alpha$+$^6$He(0$_1^+$)]$_{L=1}$. 

\begin{figure}
  \includegraphics[width=5.7cm]{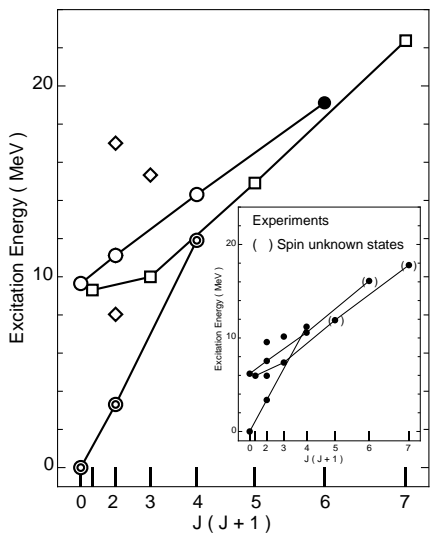} 
\vspace{-2mm}
\caption{Band structures in $^{10}$Be. 
The bands shown by the double circles and the white squares have 
($\pi_{3\slash2}^-$)$^2$ and ($\pi_{3\slash2}^-\sigma_{1\slash2}^+$), 
respectively, while that with the white circles does 
($\sigma_{1\slash2}^+$)$^2$. The solid symbols have a developed 
$\alpha$+$^6$He cluster structure. The inset is a same figure but for 
the observed states \cite{FREE,LIEN}. The spins for the states in 
parentheses are tentatively suggested in Ref.~\cite{FREE}. 
}
\label{ejj}
\end{figure}

To see band structures in $^{10}$Be, we solve GCM Eq.~(\ref{GCM}) 
for the higher spins with a natural 
parity $(-1)^J$. The calculated bands are shown 
in Fig.~\ref{ejj} with an inset showing the respective observed states 
\cite{FREE,LIEN}. 
The moment of inertia of individual bands is well reproduced. 
Furthermore, our model predicts the existence of 
the higher spin states which were suggested by a recent experiment 
\cite{FREE}. In the ($\sigma_{1\slash2}^+$)$^2$ 
band shown by the white circles, we find the enhancement of the 
[$\alpha$+$^6$He(2$_1^+$)]$_{L=4}$ state at a maximum spin
(solid circle). 
In the present calculation, all the excitation energies is higher than 
those of the observed states. This can be improved by 
optimizing the nucleon--nucleon force and the 
radius $b$ to reproduce the cluster's threshold energies. 
The reproduction of threshold 
energies is quite important for the study on the decay 
width. 
The quantitative analysis including the decay width will 
be given in forthcoming papers. 

In summary, we proposed a generalized two-center cluster model 
(GTCM) and discussed  its application to the 
$^{10}$Be=$\alpha$+$\alpha$+$n$+$n$ system with $J^\pi$=0$^+$ 
and 1$^-$. 
The adiabatic energy surfaces depending on the $\alpha$--$\alpha$ 
distance were calculated. The adiabatic eigenstates have the molecular 
orbital configuration at an internal region, while they becomes 
the cluster-coupling states at an external one. The middle distances 
correspond to the transitional region having an intermediate 
coupling-scheme between the internal regions and the external ones. 
It should be noticed that, in our model, both schemes of the 
molecular-orbitals and cluster-coupling schemes can be naturally 
described in an equal footing without any difficulty relevant to the 
double-projection procedure. 

Finally, we solved the coupling between the relative motions of 
clusters and the intrinsic motion of valence neutrons. 
The low-lying  0$^+$ states and the lowest 
1$^-$ one have the respective molecular-orbital configurations, 
which are consistent with previous studies based on 
MOM \cite{ITA} and AMD \cite{ENYO}. In addition, our model predicts 
the possible appearance of the [$\alpha$+$^6$He(2$_1^+$)]$_{L=4}$ 
cluster-structures in the lowest 6$^+$ state. 
Furthermore, we theoretically obtain the second 1$^-$ state 
with the [$\alpha$+$^6$He(0$_1^+$)]$_{L=1}$ structure and 
expect to observe it experimentally. To see correspondence with 
experiments, it is necessary to analyze resonant states above threshold 
energies. We are now going to tackle this problem. 

In conclusion, we can say that GTCM well describes 
the low-lying molecular orbitals obtained by other theoretical models 
\cite{ITA, ENYO, OGAWA}. Furthermore, this model naturally reproduce 
the asymptotic cluster-coupling states which are described by the 
traditional cluster model \cite{DES}
in an equal footing. 
These results indicate that the present GTCM is applicable to 
the study of low-lying and high-lying states in $^{10}$Be. 
Since the $\alpha$+$\alpha$+2$n$ model for $^{10}$Be is a special 
case (x=2) of a general $\alpha$+$\alpha$+x$n$ model, 
it is very easy to apply GTCM to systematic studies of resonances 
observed in excited Be-isotopes as well as their low-lying states. 
In particular, its application to $^{12}$Be = $\alpha$+$\alpha$+4$n$ 
is very interesting because of their accumulated 
experimental results \cite{FREE,SAITO}. 
A direct extension to the $\alpha$+$\alpha$+4$n$ model for $^{12}$Be 
is is now under progress. 

The authors would like to thank N.\ Itagaki, A.\ Ohnishi and 
Y.\ Sakuragi for their valuable discussions. 
One of the present authors (M.\ I.~) also would like to thank the 
Japan Society for the Promotion of Science (JSPS) for financial support. 
This work was performed as a part of the "Research Project for Study of
Unstable Nuclei from Nuclear Cluster Aspects " at RIKEN.

\end{document}